\documentclass[12pt,a4paper]{icicel}

\newtheorem{theorem}{Theorem}[section]

\usepackage[dvips]{graphicx}
\usepackage{subfigure}
\usepackage{epsfig}
\usepackage{epsf}
\usepackage{algorithm} 
\usepackage{algorithmic}

\newcommand{\bm}[1]{\mbox{\boldmath{$#1$}}}

\title[]
      { Wireless Information and Energy Transfer for Decode-and-Forward Relying  MIMO-OFDM Networks}

\author[Guanyao Du, Zhilong Dong, Ke Xiong and Zhengding Qiu]{}

\begin{document}

\maketitle

\centerline{\scshape  Guanyao Du$^1$, Zhilong Dong$^2$, Ke Xiong$^1$ and Zhengding Qiu$^1$}
 \medskip
{\footnotesize
\centerline{$^1$School of Computer and Information Technology, Beijing Jiaotong University}
\centerline{Beijing 100044, P. R. China}
\centerline{\{\, 08112076; kxiong; zdqiu\, \}@bjtu.edu.cn} }

\medskip

{\footnotesize
\centerline{$^2$State Key Lab. of Scientific and Engineering Computing Academy of Mathematics and Systems Science, }
\centerline{Chinese Academy of Sciences}
\centerline{Beijing 100190, P. R. China}
\centerline{zldong@lsec.cc.ac.cn}
\centerline{Corresponding Author: Ke Xiong} }
\medskip



\medskip

\begin{abstract}

{\em This paper investigates the system achievable rate and optimization for the multiple-input multiple-output (MIMO)-orthogonal frequency division multiplexing (OFDM) system with an energy harvesting (EH) relay. Firstly we propose a time switching-based relaying (TSR) protocol to enable the simultaneous information processing and energy harvesting at the relay. Then, we discuss its achievable rate performance theoretically and formulated an optimization problem to maximize the system achievable rate. As the problem is difficult to solve, we design an Augmented Lagrangian Penalty Function (ALPF) method for it. Extensive simulation results are provided to demonstrate the accuracy of the analytical results and the effectiveness of the ALPF method.}
\end{abstract}
\section{Introduction}
Energy harvesting (EH) has emerged as a promising approach to overcome the limited energy budget of wireless networks \cite{EH1}-\cite{OFDM-MIMO-22} in recent years. Compared with conventional EH sources (e.g., solar, wind, thermoelectric effects or other physical phenomena \cite{EH1}-\cite{EH2}), one prospective way is to harvest energy from the ambient radio-frequency (RF) signals \cite{SWIET}-\cite{OFDM-MIMO-22}, which is referred to as simultaneous wireless information and energy transfer (SWIET).

The idea of SWIET was first proposed in \cite{SWIET}, where the performance tradeoff between the energy and information
rate was studied. Later, it was widely investigated in various models \cite{OFDM}-\cite{OFDM-MIMO-22}. Specifically, in \cite{OFDM}, the multi-user orthogonal frequency division multiplexing (OFDM) system was considered, where the optimal design of SWIET was obtained. In \cite{MIMO}, a three node multiple-input multiple-output (MIMO) broadcasting system was considered, where the rate-energy bound and region were studied.

Recently, efforts have been made to apply MIMO and OFDM technologies to wireless communication system in order to support high data rates and provide high spectral efficiency. However, only a few works have investigated the MIMO-OFDM system with SWIET technology \cite{OFDM-MIMO-12}-\cite{OFDM-MIMO-22}. Specifically, in \cite{OFDM-MIMO-12} and \cite{OFDM-MIMO-22}, a two-hop MIMO-OFDM relaying system was considered, where the relay employed an amplify-and-forward (AF) cooperative scheme, and the optimum performance boundaries and the rate-energy region were investigated.

In this paper, we also focus on the SWIET for a two-hop MIMO-OFDM relaying system, where a source transmits its information to the destination with the help of an energy-constrained relay, as the relay network has great penitential by employing some advanced technologies, see e.g., \cite{Fan1,Fan2}.  The main contributions of this paper can be summarized as follows.
 Firstly, by applying the time-switching receiver architecture proposed in \cite{architecture}, we design a transmission protocol to enable the simultaneous information processing and EH at the decode-and-forward (DF) relay. Secondely, we discuss the system achievable rate performance theoretically and formulated an optimization problem to explore the system performance limit. Thirdly, as the problem is difficult to solve, we design an Augmented Lagrangian Penalty Function (ALPF) method for it. Finally, extensive simulation results are provided to demonstrate the accuracy of the analytical results and the effectiveness of the ALPF method.
\section{System Model and Protocol Description}\label{sec_model}
\subsection{Assumptions and Notations}
We consider a half-duplex two-hop DF relaying system which consists of a source $\rm S$, a destination $\rm D$, and an energy-constrained relay $\rm R$. All nodes are equipped with multiple antennas, and the number of antennas at $\rm S$, $\rm R$, and $\rm D$  are denoted by $N_{\rm S}$, $N_{\rm R}$ and $N_{\rm D}$, respectively.

$\rm S$ has fixed energy supplying and wants to transmit information to $\rm D$. We assume that the direct link between $\rm S$ and $\rm D$  is unavailable. A relay $\rm R$ is used to assist the information forwarding from $\rm S$ to $\rm D$. $\rm R$ is energy constrained since it has no internal energy source. Thus, it relies on external charging. Specifically, $\rm R$ harvests energy from the received RF signals transmitted from $\rm D$, and uses all the harvested energy to assist the information relaying. We also assume that all nodes have perfect knowledge of the channels of both hops.
Broadband communication OFDM is considered in this model, and the frequency-selective channel with total frequency band $\mathfrak{B}$ is divided into $K$  frequency-flat sub-channels. Moreover, we consider the block fading channel, where the channel gain of each sub-channels remains constant during each round of relaying transmission.

\subsection{The Proposed Time Switching-based Relaying (TSR) Protocol}
\begin{figure}
\centering
\includegraphics[width=0.6\textwidth]{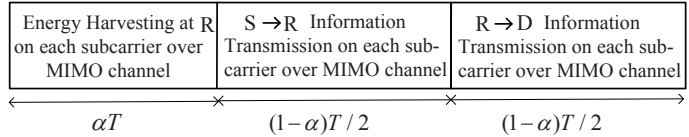}
\caption{System model and parameters}
\label{figsystemmodel}
\end{figure}

Fig.~\ref{figsystemmodel} depicts the main transmission process in the proposed TSR protocol. By considering the time switching receiver architecture described in \cite{architecture}, the proposed TSR protocol consists of three phases: the energy transfer phase, the information transmission from  $\rm S$ phase and the information relaying from $\rm R$  phase, as shown in Fig.~\ref{figsystemmodel}. The time durations assigned to each phase are  $\alpha T$, $(1-\alpha)T/2$  and $(1-\alpha)T/2$, respectively, where $0\leq\alpha\leq1$  denotes the time assignment factor.

In the first phase, $\rm S$  transfers energy to $\rm R$, and the received signal at $\rm R$  for energy harvesting is given as $\mathbf y^{\rm {(EH)}}_{{\rm R},k}=\mathbf H_{1,k}\mathbf x_k+\mathbf n_{{\rm R},k}$,
where $\mathbf H_{1,k}\in\mathbb{C}^{N_{\rm R}\times N_{\rm S}}$  denotes the channel matrix from $\rm S$ to $\rm R$ at hop 1 over the $k$-th subcarrier,  $\mathbf x_k$ denotes the transmitted signal vector for energy transfer over subcarrier $k$, and $\mathbf n_{{\rm R},k} \sim CN(0,\sigma _{\rm{R}}^2\mathbf I_{{N_{\rm{R}}}})$ is a $N_{\rm R}\times 1$  additive white Gaussian noise (AWGN) vector at $\rm R$.

Thus, the harvested energy at $\rm R$ over subcarrier  $k$ is $E_{{\rm R},k}=\alpha \eta {\left\| \mathbf H_{1,k}\mathbf x_k \right\| ^2}$, where $0<\eta\leq1$  denotes the energy conversion efficiency. The total harvested energy over $K$  subcarriers can be given by $E = \sum_{k = 1}^K {{E_{{\rm{R}},k}}}$, and note that, the total transferred energy is limited by the available power at $\rm S$, i.e. $\sum_{k = 1}^K {\rm {tr}(\mathbf x_k} \mathbf x_k^H) \le {\mathcal{P}_{\rm{s}}}$. Since all the harvested energy in the first phase is used to relay the information in the third phase, the available transmit power at $\rm R$ in the information relaying phase is given by
\begin{small}
\begin{equation}
{\mathcal{P}_{\rm{R}}} = \frac{E}{{(1 - \alpha )T/2}} = \frac{{2\alpha \eta }}{{1 - \alpha }}\sum_{k = 1}^K {{{\left\| \mathbf H_{1,k}\mathbf x_k  \right\|}^2}}
\end{equation}
\end{small}
In the second phase, i.e. the information transmission from $\rm S$ phase, $\rm S$  delivers the signal vector $\mathbf{s}_k\in\mathbb{C}^{N_{\rm S}\times 1}$  to $\rm R$, and the received signal at $\rm R$ over the $k$-th subcarrier can be represented as
\begin{equation}
\mathbf y_{{\rm R},k}=\mathbf W_{{\rm R},k}\mathbf H_{1,k} \mathbf F_{{\rm S},k} \mathbf{s}_k+\mathbf n_{{\rm R},k}\label{sk}
\end{equation}
where ${\rm{E[}}{\mathbf{s}_k}\mathbf{s}_k^H] = {\mathbf I_{{N_{\rm{S}}}}}$, $\mathbf F_{{\rm S},k}\in\mathbb{C}^{N_{\rm S}\times N_{\rm S}}$  denotes the precoding matrix at $\rm S$,  and $\mathbf W_{{\rm R},k}\in\mathbb{C}^{N_{\rm R}\times N_{\rm R}}$  is the receiver filter deployed at $\rm R$ to detect the relaying signal. Then, the achievable rate at $\rm R$ is given by
\begin{equation}
\mathbf R_{{\rm{R}},k}={\rm{log}}_2 \left| {\mathbf I + \mathbf W_{{\rm R},k} \mathbf H_{1,k} \mathbf F_{{\rm S},k}\mathbf F_{{\rm S},k}^H \mathbf H_{1,k}^H \mathbf W_{{\rm{R}},k}^H \sigma _{\rm{R}}^{ - 2}} \right|  \label{R_r}
\end{equation}

In the third phase, i.e. the information relaying from $\rm R$  phase,  $\rm R$ decodes the signal $\mathbf{s}_k$  from \eqref{sk} and forwards it to $\rm D$  by multiplying a forwarding matrix $\mathbf F_{{\rm R},m}\in\mathbb{C}^{N_{\rm R}\times N_{\rm R}}$. By considering the subcarrier pairing, we assume that the $k$-th subcarrier over hop-1 in the second phase is paired with the $m$-th subcarrier over hop-2 in the third phase, and call them subcarrier pair (SP) $(k,m)$. Thus, the received signal at $\rm D$  over the SP $(k,m)$  can be expressed as
\begin{equation}
\mathbf y_{{\rm D},m}=\mathbf W_{{\rm D},m}\mathbf H_{2,m} \mathbf F_{{\rm R},m} \mathbf{s}_k+\mathbf n_{{\rm D},m}
\end{equation}
where   $\mathbf H_{2,m}\in\mathbb{C}^{N_{\rm D}\times N_{\rm R}}$ denotes the channel matrix from $\rm R$  to $\rm D$  at hop 2 over the $m$-th subcarrier, and  $\mathbf W_{{\rm D},m}\in\mathbb{C}^{N_{\rm D}\times N_{\rm D}}$ is the receiver filter deployed at $\rm D$  to detect the relaying signal. $\mathbf n_{{\rm D},m} \sim CN(0,\sigma _{\rm{D}}^2\mathbf I_{{N_{\rm{D}}}})$   is the $N_{\rm D}\times 1$  AWGN vector at $\rm D$. Then, the achievable rate at $\rm D$ is given by
\begin{equation}
\mathbf R_{{\rm{D}},m}={\rm{log}}_2 \left| {\mathbf I + \mathbf W_{{\rm D},m} \mathbf H_{2,m} \mathbf F_{{\rm R},m}\mathbf F_{{\rm R},m}^H \mathbf H_{2,m}^H \mathbf W_{{\rm{D}},m}^H \sigma _{\rm{D}}^{ - 2}} \right|   \label{R_d}
\end{equation}

Since the achievable rate for the two-hop relaying system is bounded by the minimum of \eqref{R_r} and \eqref{R_d}, the achievable rate over the SP  $(k,m)$ is given by
\begin{equation}
\mathbf R_{k,m}=\frac{\mathfrak{B}}{K}\cdot\frac{1-\alpha}{2}{\rm{min}} (\mathbf R_{{\rm{R}},k},\mathbf R_{{\rm{D}},m})  \label{R_min}
\end{equation}
where $\mathfrak{B}$  denotes the total bandwidth of the OFDM system and $(1-\alpha)/2$  results from the transmission duty cycle loss in TSR protocol for the two-hop relaying system.
\section{Achievable Rate Analysis and Optimization Problem Formulation}\label{sec_analysis}
By performing the singular value decomposition (SVD) on $\mathbf H_{1,k}$  and $\mathbf H_{2,m}$, the MIMO channels of the two-hop relaying system can be decomposed into multiple parallel independent subchannels with different gain. Specifically, the SVD of the channel matrices is given by $\mathbf H_{i,q}=\mathbf U_{i,q}\mathbf \Lambda_{i,q}\mathbf V_{i,q}^H  \label{SVD}$,
where $q=k$  for  $i=1$, and $q=m$  for $i=2$. Both  $\mathbf U_{i,q}$ and $\mathbf V_{i,q}^H$  are unitary, and  $\mathbf \Lambda_{i,q}\in\mathbb{C}^{{\rm{Rank}(\mathbf H _{\emph{i,q}})}\times {\rm {Rank}(\mathbf H_\emph{i,q})}}$ is a diagonal matrix whose diagonal elements $\{ \sqrt {{\lambda _{i,l}}} \} _{l = 1}^{{\rm{Rank}}({\mathbf H_{i,q}})}$  are nonzero singular values of $\mathbf H_{i,q}$  in descending order.

Due to the full channel state information (CSI) at all the nodes, we can use the SVD of the channel matrices to determine the precoding matrix and receiver filter matrices at the transmitter and receiver. Specifically, we choose the precoding matrix at $\rm S$, the forwarding matrix at $\rm R$ and the receiver filters deployed at $\rm R$ and $\rm D$  as $\mathbf F_{{\rm{S}},k}=\sqrt{P_{{\rm{S}},k}}\mathbf V_{1,k}$, $\mathbf F_{{\rm{R}},m}=\sqrt{P_{{\rm{R}},m}}\mathbf V_{2,m}$, $\mathbf W_{{\rm{R}},k}=\mathbf U_{1,k}^H$  and  $\mathbf W_{{\rm{D}},m}=\mathbf U_{2,m}^H$, respectively, where $P_{{\rm{S}},k}$ and  $P_{{\rm{R}},m}$ denote the available transmit power at $\rm S$ and $\rm R$, respectively.

Substituting \eqref{SVD} and above designed matrices into \eqref{R_min}, the achievable rate over the SP $(k,m)$  can be rewritten as
\begin{equation}
\mathbf R_{k,m}=\frac{(1-\alpha)\mathfrak{B}}{2K}{\rm{min}} ({\rm{log}}_2 \left| \mathbf I + P_{{\rm{S}},k}\mathbf \Lambda_{1,k} \mathbf \Lambda_{1,k}^H \sigma _{\rm{R}}^{ - 2} \right|, {\rm{log}}_2 \left|\mathbf I + P_{{\rm{R}},m}\mathbf \Lambda_{2,m} \mathbf \Lambda_{2,m}^H \sigma _{\rm{D}}^{ - 2} \right|)
\end{equation}

Though the above mentioned operations, each SP $(k,m)$  is divided into $N$  available end-to-end (E2E) subchannels, where $N$ denotes the number of available spatial subchannels per OFDM subcarrier over the two hops, which is bounded to the minimum number of spatial subchannels of each hop, i.e., $N=\rm{min}\{{\rm{Rank}(\mathbf H _{1,\emph{k}})},{\rm{Rank}(\mathbf H _{2,\emph{m}})}\}=\rm{min}\{ \emph{N}_{\rm S},\emph{N}_{\rm R},\emph{N}_{\rm D}\}$. Since there are $K$  subcarriers, the total number of effective E2E subchannels in the MIMO-OFDM system is $KN$. We introduce the subscript $n \buildrel \Delta \over = (k - 1)K + l$ and $n' \buildrel \Delta \over = (m- 1)K + l'$  to simplify the notation, where $1\leq l,l'\leq N$. As a result, $1\leq n,n'\leq KN$, and the above mentioned SP $(k,m)$   can be rewritten as SP $(n,n')$   which means that the  $n$-th subchannel over hop 1 is paired with the  $n'$-th subchannel over hop 2. Further, we define $P_{{\rm S},n} \buildrel \Delta \over =\mathcal{P}_{\rm S}\mu_n$,  $P_{{\rm R},n'} \buildrel \Delta \over =\mathcal{P}_{\rm R}\overline\mu_{n'}$, where $\mu_n$ and  $\overline\mu_{n'}$ denote the power allocating factor at $\rm S$ for subchannel $n$  over hop-1 and the power allocating factor at $\rm R$  for subchannel  $n'$ over hop-2, respectively. Consequently, the achievable rate $ R_{n,n'}$  over SP  $(n,n')$  can be expressed as
\begin{equation}
 R_{n,n'}=\frac{(1-\alpha)\mathfrak{B}}{2K}{\textmd{min}} (\textmd {log}_2 (1 + \frac{\mathcal{P}_{\rm S}\mu_n \lambda _{1,n}}{\sigma_{\rm R}^2}),\textmd{log}_2 (1 + \frac{\mathcal{P}_{\rm R}\overline\mu_{n'} \lambda _{2,n'}}{\sigma_{\rm D}^2}))
\end{equation}

Thus, the achievable rate of the TSR protocol in the DF MIMO-OFDM relaying system is given by $C^{(\rm TSR)}=\sum _{n = 1}^{KN} \sum _{n' = 1}^{KN}\theta_{n,n'}  R_{n,n'}$, where $\theta_{n,n'}\in\{0,1\}$  denotes the subchannel-paring, and the optimization problem of maximizing the achievable rate for a DF MIMO-OFDM relaying system can be formulated as follows
\begin{small}
\begin{equation}\label{Problem}
\begin{aligned}
\mathop {\max }\limits_{\mathbf X_{\rm S},\mu_n, \overline{\mu}_{n'},\theta_{n,n'}, \alpha}{\kern 1pt} {\kern 1pt}{\kern 1pt} {\kern 1pt}& C^{(\rm TSR)}\\
s.t.\ &\sum_{n=1}^{KN}\mu_n\leq 1,{\kern 3pt}\sum_{n'=1}^{KN}\overline\mu_{n'}\leq 1,{\kern 4pt}\mu_n\geq0,{\kern 4pt}\mu_{n'}\geq0\\
&{\kern 4pt}\sum \limits_{k = 1}^{K} \textmd{tr}(\mathbf x_k \mathbf x_k^H)\leq \mathcal{P}_{\rm S}, {\kern 4pt}\mathbf X_i\succeq0\\
&\sum \limits_{n = 1}^{KN} \theta_{n,n'}=1, {\kern 3pt}\sum \limits_{n' = 1}^{KN} \theta_{n,n'}=1,{\kern 3pt}\theta_{n,n'}\in\{0,1\},{\kern 4pt}0\leq\alpha\leq1
\end{aligned}
\end{equation}
\end{small}

Specifically, $\theta_{n,n'}=1$  means that the  $n$-th subchannel over hop-1 is paired with the  $n'$-th subchannel over hop-2. Otherwise, $\theta_{n,n'}=0$. Let  $\mathbf X_{k}=E\{\mathbf x_k \mathbf x_k^H\}$ denote the covariance matrix of $\mathbf x_k$, $\mathbf X_{\rm S}=\{\mathbf X_1,\mathbf X_2,...,\mathbf X_{k}\}$  indicates the energy transfer pattern at $\rm S$.
\section{Achievable Rate Optimization}\label{sec_optimization}
The main ideas to solve \eqref{Problem} are as follows:
   Firstly, only energy is delivered in the first phase, which means that only $\textit{\textbf{X}}_{\rm S}$  needs to be optimized and it is independent with other variables. Thus, we could design  $\textit{\textbf{X}}_{\rm S}$ independently which will not affect the global optimality.
 Secondly, according to the separation principle designed in \cite{seperated_princeple}, the joint channel pairing and power allocation optimization problem can be decoupled into two separate sub-problems. So, we can optimize  $\theta$ independently without considering other variables.
 Thirdly, based on the optimal $\textit{\textbf{X}}_{\rm S}^{\sharp}$  and  ${\bm \theta^{\sharp}}$, we propose an Augmented Lagrangian Penalty Function (ALPF) method to jointly optimize $\mu_n$, $\overline{\mu}_{n'}$ and $\alpha$ to maximize $C^{\rm (TSR)}$.

\subsection{Optimal $\textit{\textbf{X}}_{\rm S}^{\sharp}$ and optimal ${\bm \theta^{\sharp}}$  for TSR}
To achieve the maximum energy transfer, all power at $\rm S$ for energy delivery should be allocated to the subcarrier with the maximum  $\|\widetilde{\mathbf h}_{{\rm S},1}^{(i)}\|^2$, where $i\in\{1,2,...,K\}$, and $\widetilde{\mathbf h}_{{\rm S},1}^{(i)}$ denotes the first column of matric $\mathbf H_{{\rm S},i} \mathbf V_{{\rm S},i}$.

\begin{theorem}
The optimal subchannel pairing ${\bm \theta^{\sharp}}$ is performed in the order of sorted channel gain which means that the subchannel with $i$-th largest channel gain (normalized against the noise power) over hop-1 should be paired with the subchannel with $i$-th largest channel gain (also normalized against the noise power) over hop-2.
\end{theorem}
{\bf Proof:}\ According to the separation principle in \cite{seperated_princeple}, the joint channel pairing and power allocation optimization problem can be operated in a separated manner, and the optimal channel pairing is performed individually at the relay in the order of sorted channel gain.
\subsection{Joint optimal $\mu_n$, $\overline{\mu}_{n'}$ and $\alpha$  for TSR}
According to the max-flow min-cut theorem, the achievable rate of each subchannel is limited by the minimal rate over the two hops. Thus, the rates of two hops are equal to each other when $C^{\rm (TSR)}$  is maximized. Further, by substituting the optimal  $\textit{\textbf{X}}_{\rm S}^{\sharp}$ and ${\bm \theta^{\sharp}}$  obtained from subsection 4.1 into \eqref{Problem}, the original optimization problem can be rewritten as follows
\begin{small}
\begin{equation}\label{Standard}
\begin{aligned}
\mathop {\min }\limits_{\mu_n, \overline{\mu}_{n'}, \alpha}{\kern 1pt} {\kern 1pt}{\kern 1pt} {\kern 1pt}& \frac{{(\alpha-1)\mathfrak{B}}}{2K}\sum_{n=1}^{KN}{\rm{log}}_2(1 + \frac{\mathcal{P}_{\rm S}\mu_n \lambda _{1,n}}{\sigma_{\rm R}^2})\\
s.t.\ &\sum_{n=1}^{KN}\mu_n\leq 1,{\kern 4pt}\sum_{n'=1}^{KN}\overline\mu_{n'}\leq 1,{\kern 3pt}\mu_n\geq0,{\kern 3pt}\mu_{n'}\geq0\\
&\frac{\mathcal{P}_{\rm S}\mu_n \lambda _{1,n}}{\sigma_{\rm R}^2}=\frac{\mathcal{P}_{\rm R}\overline\mu_{n'} \lambda _{2,n'}}{\sigma_{\rm D}^2}, {\kern 4pt}\text{for} {\kern 4pt}\theta_{n,n'}=1,{\kern 4pt}0\leq\alpha\leq1
\end{aligned}
\end{equation}
\end{small}
It can be observed that the problem in \eqref{Standard} is still a nonlinear and non-convex optimization problem which is difficult to solve. Here, we introduce an Augmented Lagrangian Penalty Function (ALPF) method \cite{ALPF1}-\cite{ALPF2} to find the joint optimal $\mu_n$, $\overline{\mu}_{n'}$ and $\alpha$.


The unconstrained augmented Lagrangian penalty function of \eqref{Standard} can be written as
\begin {small}
\begin{equation}\label{Lagrangian}
\begin{aligned}
  P(\textit{\textbf{x}},\nu,\sigma)=\frac{(\alpha-1)\mathfrak{B}}{2K}\sum_{n=1}^{KN}\rm{log}_2(1+A_n\mu_n)-\nu_1 (\sum_{n=1}^{KN}\mu_n+S_1-1)-\nu_2(\sum_{n'=1}^{KN}\overline{\mu}_{n'}+S_2-1)\\
  -\sum_{(n,n')}\nu_{n,n'}(A_n\mu_n-\frac{2\alpha}{1-\alpha}B_{n'} \overline{\mu}_{n'})+
  \frac{1}{2}\sigma_1(\sum_{n=1}^{KN}\mu_n+S_1-1)^2+
  \frac{1}{2}\sigma_2(\sum_{n'=1}^{KN}\overline{\mu}_{n'}+S_2-1)^2\\
  +\sum_{(n,n')}\frac{1}{2}\sigma_{n,n'}(A_n\mu_n-\frac{2\alpha}{1-\alpha}B_{n'}\overline{\mu}_{n'})^2
\end{aligned}
\end{equation}
\end {small}
where $A_n=\frac{\mathcal{P}_{\rm S} \lambda _{1,n}}{\sigma_{\rm R}^2}$, $B_{n'}=\frac{\mathcal{P}_{\rm R} \lambda _{2,n'}}{\sigma_{\rm D}^2}$, $\textit{\textbf{x}}=(\alpha,\mu_n,\overline{\mu}_{n'},S_1,S_2)$, $S_1$ and $S_2$ are positive slack variables, $\nu$ and $\sigma$ denote the Lagrangian multipliers and the penalty parameter, respectively. We also define the constraint violation function as $C^{(-)}(\textit{\textbf{x}})=(\sum_{n=1}^{KN}\mu_n+S_1-1,\sum_{n'=1}^{KN}\overline{\mu}_{n'}+S_2-1, A_n\mu_n-\frac{2\alpha}{1-\alpha}B_{n'}\overline{\mu}_{n'})$

In the $k$-th iteration, with $\nu_i^{(k)}$ and $\sigma_i^{(k)}$, $\textit{\textbf{x}}^{(k+1)}$ can be obtained by
\begin{equation}\label{subproblem}
  \textit{\textbf{x}}^{(k+1)}=\arg\min P(\textit{\textbf{x}}^{(k)},\nu^{(k)},\sigma^{(k)})
\end{equation}
Then the lagrange multipliers can be updated as $\nu_1^{(k+1)}=\nu_1^{(k)}-\sigma_1^{(k)}(\sum_{n=1}^{KN}\mu_n^{(k+1)}+S_1^{(k+1)}-1)$, $\nu_2^{(k+1)}=\nu_2^{(k)}-\sigma_2^{(k)}(\sum_{n'=1}^{KN}\overline{\mu}_{n'}^{(k+1)}+S_2^{(k+1)}-1)$ and $\nu_{n,n'}^{(k+1)}=\nu_{n,n'}^{(k)}-\sigma_{n,n'}^{(k)}(A_n\mu_{n}^{(k+1)}-\frac{2\alpha^{(k+1)}}{1-\alpha^{(k+1)}} B_{n'}^{(k+1)}\overline{\mu}_{n'}^{(k+1)})$.
And the penalty parameters can be updated by
\begin {small}
\begin{equation}\label{penalty1}
\sigma_1^{(k+1)}=
\begin{cases}
\sigma_1^{(k)}, &\text{if}\ |\sum_{n=1}^{KN}\mu_n^{(k+1)}+S_1^{(k+1)}-1|\le\frac{1}{4}|\sum_{n=1}^{KN}\mu_n^{(k)}+S_1^{(k)}-1|,\\
\max\{10\sigma_1^{(k)},k^2\}, &\text{otherwise.}
\end{cases}
\end{equation}
\begin{equation}\label{penalty2}
\sigma_2^{(k+1)}=
\begin{cases}
\sigma_2^{(k)}, &\text{if}\ |\sum\overline{\mu}_{n'}^{(k+1)}+S_2^{(k+1)}-1|\le\frac{1}{4}|\sum\overline{\mu}_{n'}^{(k)}+S_2^{(k)}-1|,\\
\max\{10\sigma_2^{(k)},k^2\}, &\text{otherwise.}
\end{cases}
\end{equation}
\begin{equation}\label{penalty3}
\sigma_{l,l'}^{(k+1)}=
\begin{cases}
\sigma_{n,n'}^{(k)}, &\text{if}\ |(A_n\mu_n^{(k+1)}-\frac{2\alpha^{(k+1)}}{1-\alpha^{(k+1)}}B_{n'}^{(k+1)}\overline{\mu}_{n'}^{(k+1)})|\le\frac{1}{4}|(A_n\mu_n^{(k)}-\frac{2\alpha^{(k)}}{1-\alpha^{(k)}} B_{n'}^{(k)}\overline{\mu}_{n'}^{(k)})|,\\
\max\{10\sigma_{n,n'}^{(k)},k^2\}, &\text{otherwise.}
\end{cases}
\end{equation}
\end {small}
We present the main steps of ALPF method as shown in Algorithm 1.

\begin{algorithm}[H] \label{ALPF}
\caption{ALPF Algorithm} 
\begin{algorithmic}[1] 
\STATE Initialize $\textit{\textbf{x}}^{(0)}$, $\nu_i^{(0)}$ and $\sigma_i^{(0)}$. $k=0$ denotes the number of iterations.
\STATE Solve \eqref{subproblem} through the projected-gradient method to obtain $\textit{\textbf{x}}^{(k+1)}$.\\
If $\|C^{(-)}(\textit{\textbf{x}}^{(k+1)})\|_{\infty}\le\epsilon$, algorithm ends. Otherwise, go to Step 3.\\
\STATE Update the penalty parameters in terms of \eqref{penalty1}-\eqref{penalty3}.\\
\STATE Update the Lagrange multipliers.\\
$k=k+1$, and return to step 2.\\
\end{algorithmic}
\end{algorithm}

\begin{figure}
\centering
\subfigure[ ]{\label{fig_rate_1a} 
\includegraphics[width=0.41\textwidth]{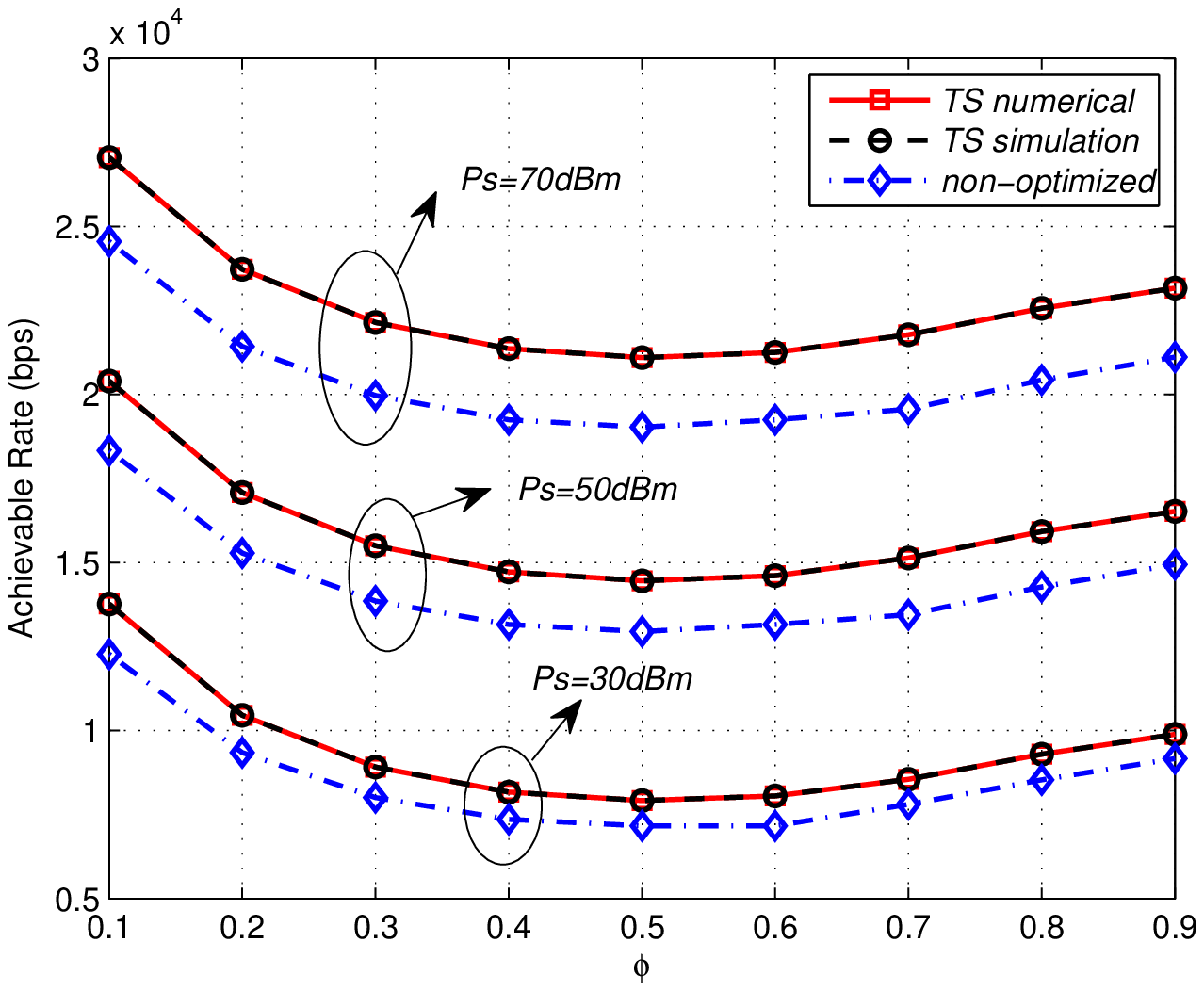}}
\subfigure[ ]{\label{fig_alpha_1b} 
\includegraphics[width=0.41\textwidth]{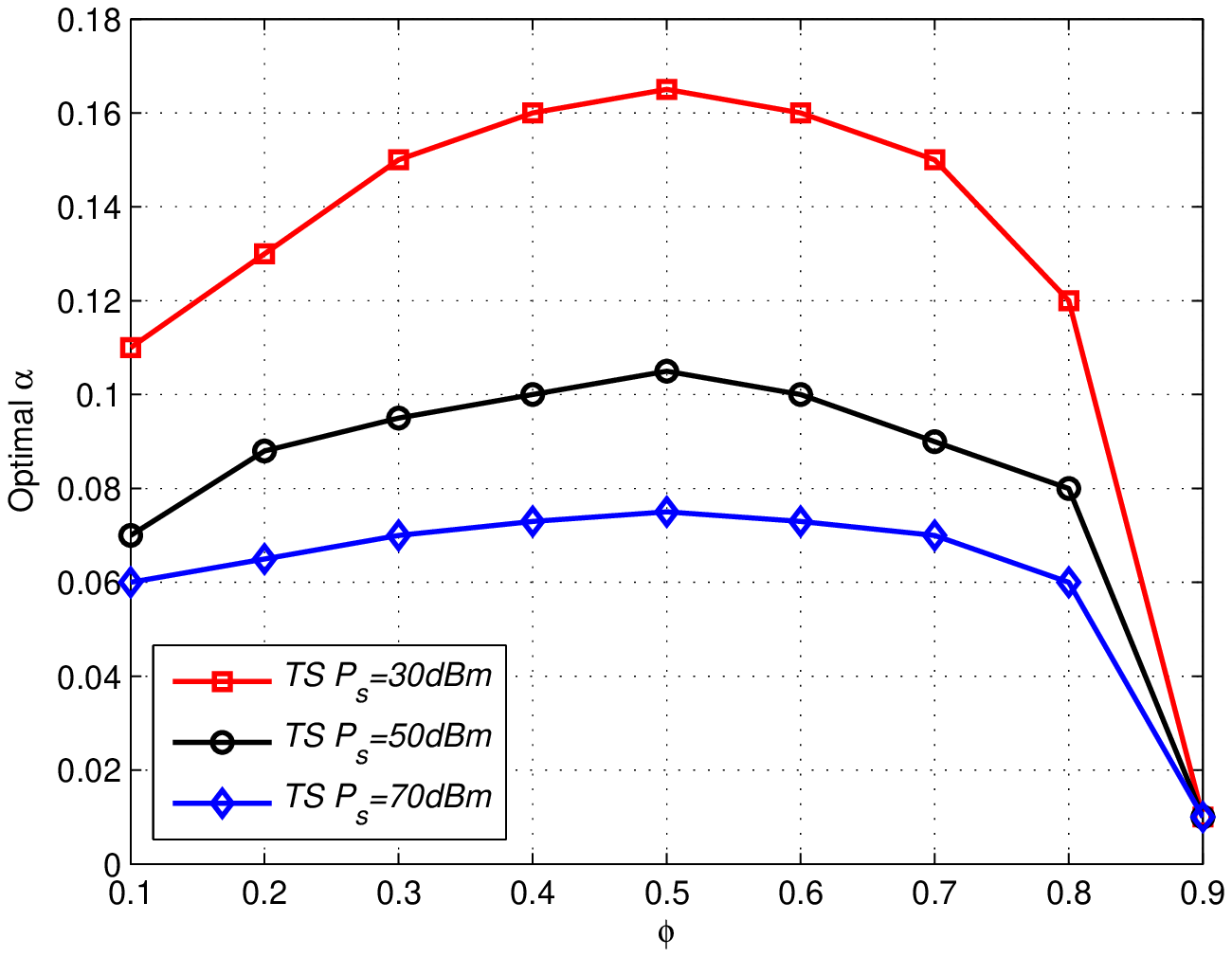}}
\caption{ (a) Optimal system achievable rate: numerical vs simulation (b) Optimal $\alpha$ vs $\phi$}
\end{figure}

\section{Numerical Example}\label{sec_simulation}
In this section, we provide some numerical results.
The distance between $\rm S$ and $\rm D$, which is denoted as $d_{\rm {SD}}$, is used to be the reference distance, and the path loss factor is set to be 4. The variable $\phi\in(0,1)$ denotes the ratio of the distance between $\rm S$ and $\rm R$, i.e. $d_{\rm {SR}}=\phi d_{\rm {SD}}$. Unless specifically stated, we set $\eta=1$, and the total system bandwith is set to be $\mathfrak{B}=1$kHz, so that each subcarrier is allocated with $1/K$kHz. The total receiving noise at $\rm R$ and $\rm D$ over the total bandwidth is set to be $10^{-6}$W, so that the noise over each subchannel is set to be $\sigma_{\rm R}^2=\sigma_{\rm D}^2=\frac{10^{-6}}{K}$W. In Algorithm 1, the initial parameters are set as $\textit{\textbf{x}}^{(0)}=(0.5,1/KN,1/KN,0.05,0.05)$, $\nu_i^{(0)}=0$ and $\sigma_i^{(0)}=1$.

To show the performance gain of the optimized TSR, we also show the results of a non-optimized scheme as a benchmark. In the non-optimized scheme, $\alpha$ is set to be 0.5,  $\mu_n$ and $\overline{\mu}_{n'}$ are set as $\mu_n=\frac{\lambda _{1,n}}{\sum_{n=1}^{KN}\lambda _{1,n}}$ and $\overline{\mu}_{n'}=\frac{\lambda _{1,n'}}{\sum_{n'=1}^{KN}\lambda _{1,n'}}$, which is proportional to the singular value of each sub-channel.

Fig.~\ref{fig_rate_1a} verifies our theoretical analysis and the proposed ALPF algorithm. It can be seen that, for different available power $\mathcal{P}_{\rm S}$ at $\rm S$, the simulation results closely match with the numerical results. Moreover, the system achievable rate achieves higher with the increment of $\mathcal{P}_{\rm S}$ due to the fact that high $\mathcal{P}_{\rm S}$ leads to high SNR for the system. Fig.~\ref{fig_rate_1a} also shows the effect of relay location on the system achievable rate. Specifically, as $\phi$ increases, the achievable rates first decrease and then increase, and achieve the minimum when the relay is deployed in the middle of $\rm S$ and $\rm D$. Fig.~\ref{fig_alpha_1b} shows the optimal $\alpha$ versus $\phi$. It can be observed that, as  $\phi$ increases, the optimal $\alpha$ first increases and then decreases. This is due to the fact that, when $\rm R$ is far away from $\rm S$, the energy harvesting efficiency becomes lower, $\rm R$ needs higher $\alpha$ to collect enough energy to decode the information from $\rm S$. And when $\rm R$ is close to $\rm D$, the channel quality of $\rm R$-$\rm D$ link gets better, $\rm R$ needs less energy to relay the information for $\rm D$, which makes $\alpha$ get lower.

\begin{figure}
\centering
\includegraphics[width=0.42\textwidth]{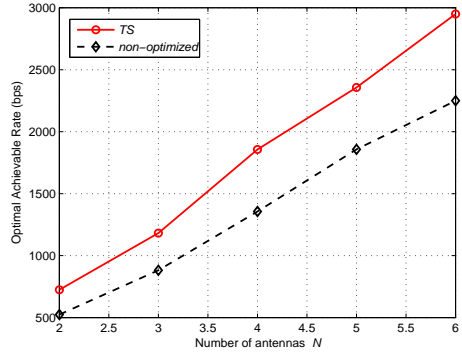}
\caption{System performance vs the number of antennas $N$}
\label{effct_n}
\end{figure}

Fig.~\ref{effct_n} shows the effect of the number of antenna $N$ on the system achievable rate. In the simulations, $K$ is set to be 2, and $N$ increases from 2 to 6. It can be observed that the system achievable rate increases as the number of antennas increases. This is due to the fact that more antennas yields more spacial subchannels, thus higher multiplex gain over subchannels can be achieved.

\section {Conclusions}\label{sec_conlusions}

In this paper, we investigated the DF MIMO-OFDM relaying system with an EH relay. We first proposed a TSR protocol for the MIMO-OFDM relaying system. Then, in order to explore the system performance limit, we formulated an optimization problem to maximize the system achievable rate, and we also proposed an ALPF method to solve it. Moreover, the effects of the relay location and the number of antennas on the system  performance were also discussed. Numerical results verify our theoretical analysis on the system achievable rate and the effectiveness of our proposed ALPF method. In the future, the performance of SWIET in  the DF MIMO-OFDM relaying system will be discussed for the case that the CSI is imperfect.



\end{document}